\documentclass[aps,prd,twocolumn,showpacs,superscriptaddress,letterpaper,amsmath,amssymb]{revtex4} 

\usepackage{graphicx}  
\usepackage{dcolumn}   
\usepackage{bm}        
\usepackage{amssymb}   
\usepackage{feynmf}
\usepackage{slashed}
\def\gsim{\lower0.5ex\hbox{$\:\buildrel >\over\sim\:$}}
\def\lsim{\lower0.5ex\hbox{$\:\buildrel <\over\sim\:$}}

\newcommand{\be}{\begin{equation}}
\newcommand{\ee}{\end{equation}}
\newcommand{\bea}{\begin{eqnarray}}
\newcommand{\eea}{\end{eqnarray}}

\newcommand{\nbox}{{\,\lower0.9pt\vbox{\hrule \hbox{\vrule height 0.2 cm
\hskip 0.2 cm \vrule height 0.2 cm}\hrule}\,}}

\begin{document}

\thispagestyle{empty}
\vspace*{-3.5cm}

\vspace{0.5in}

\title{Limits on Excited Quarks from the ATLAS Multi-jet Search}

\begin{center}
\begin{abstract}
  We interpret the recent ATLAS multi-jet search results using 20.3 fb$^{-1}$ data at $\sqrt{s}=$ 8 TeV in the context of searching for excited quarks. Within the effective field theory framework, using the null results of that search, our analysis shows that the excited quark masses below 5~TeV can be excluded at the 95\% confidence level. Our analysis also indicates that when the validation of effective field theory is considered, the limit can be largely compromised.
\end{abstract}
\end{center}
\author{Ning Zhou}
\affiliation{Department of Physics, Tsinghua University, Beijing, China}
\affiliation{Center for High Energy Physics, Tsinghua University, Beijing, China}
\affiliation{Collaborative Innovation Center of Quantum Matter, Beijing, China}
\author{Kechen Wang}
\affiliation{Center for Future High Energy Physics, Institue of High Energy Physics,\\
Chinese Academy of Sciences, Beijing, China}
\author{Kuhan Wang}
\affiliation{Department of Physics, McGill Univerity, Canada}
\maketitle

Although the Standard Model (SM) of particle physics is successful in describing many phenomena,
there are still some questions remaining unanswered. This suggests that SM might be an effective, low-energy approximation of more fundamental theory.
Quark compositeness models\cite{Pati:1974,Pati:1975,Eichten:1983,Cabibbo:1984,Hagiwara:1985,Baur:1987,Baur:1990,Redi:2013} have been proposed to reduce the number of fundamental matter consituents and explain the generational structure and the mass hierarchy of quarks. If quarks are made of constituents, then at the scale of consitituent binding energy $\Lambda$, the strong forces binding quark consituents induce flavor-diagonal Contact Interactions (CI) which can be described within an effective field theory (EFT) framework. Considering an effective four-fermion Lagrangian, at the energy much below the $\Lambda$ scale, these interactions will be suppressed by the inverse powers of $\Lambda$.
The discovery of excited quarks sharing quantum number with the SM quarks can be a strong evidence of such compositeness scenario. In our analysis, we consider the Lagrangian where an excited quark couples to an ordinary quark and a gauge boson via the gauge interactions \cite{Baur:1987,Baur:1990,Bhattacharya:2009xg}:
\begin{equation}
  \begin{split}
  \mathcal{L}_{int} = \frac{1}{2\Lambda}\bar{q}^{*}_{R}\sigma^{\mu\nu}[g_{s}f_{s}\frac{\lambda_{a}}{2}G^{a}_{\mu\nu}+gf\frac{\tau}{2}W_{\mu\nu}+g'f'\frac{Y}{2}B_{\mu\nu}]q_{L} \\
  + h.c.
  \end{split}
\end{equation}
Here $\Lambda$ denotes the compositeness scale, which is the typical energy scale of these interactions. $\sigma_{\mu\nu}$ is the Pauli spin matrix. $q^{*}_{R}$ is the excited quark field; $q_{L}$ is the quark field. $G^{a}_{\mu\nu}$, $W_{\mu\nu}$ and $B_{\mu\nu}$ are the field-strength tensors of the SU(3), SU(2) and U(1) gauge fields. $g_{s}$, $g$, $g'$ are gauge coupling constants, while $\lambda_{a}$, $\tau$, $Y$ are the coresponding gauge structure constants. The unknown dimensionless constants $f_{s}$, $f$, and $f'$, determined by the compositeness dynamics, represent the strengths of the excited quark couplings to the SM partners, which are usually assumed to be of order unity. Many searches for excited quarks have been performed in various decay channels~\cite{pdg:2012,hera:2009,d0:2004,cdf:2009,cms:2013,atlas:2014,atlas:2015}, but no evidence of their existence has been found to date.

\begin{figure}[htbp]
  \center
  \includegraphics[width=0.6\linewidth]{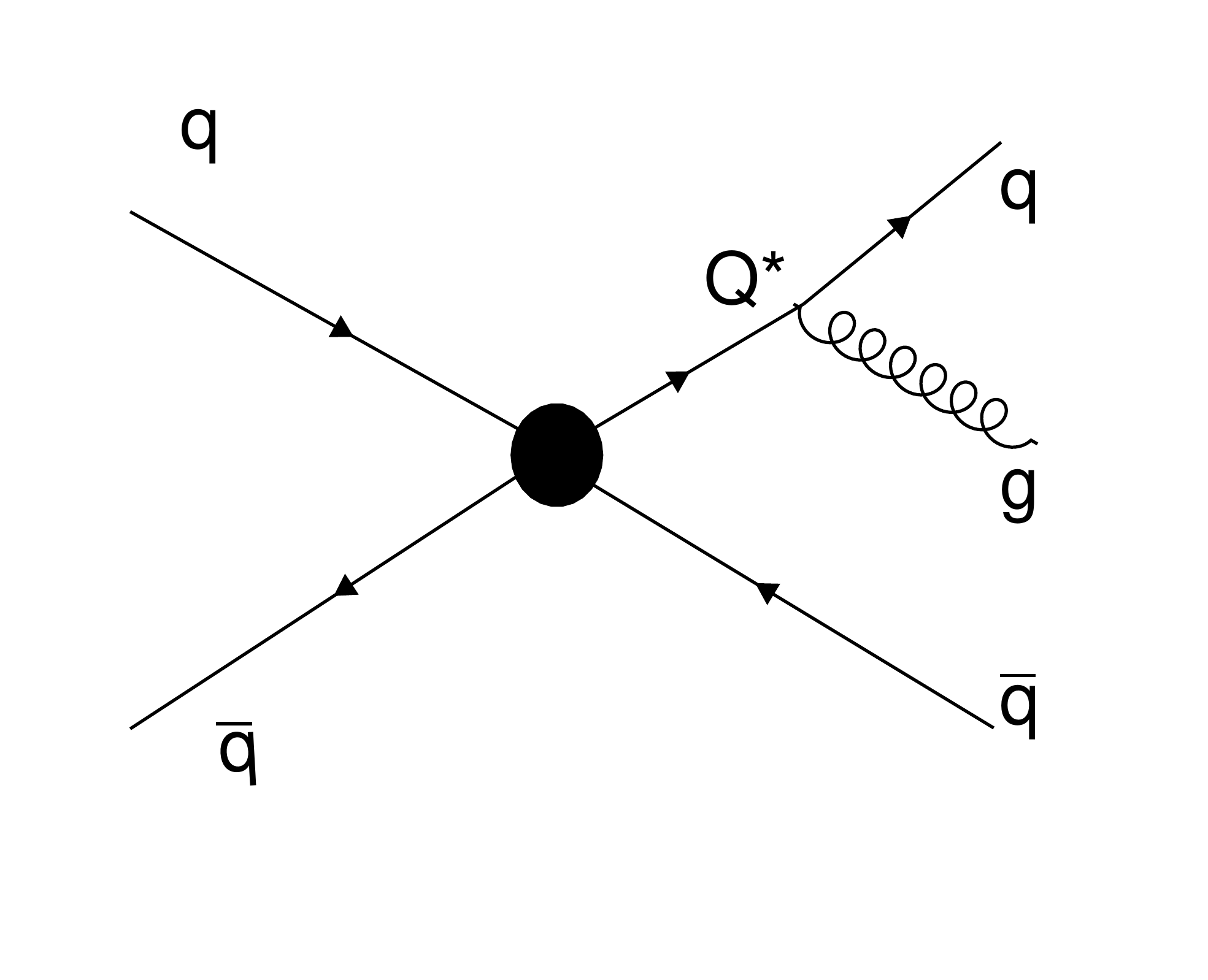}
  \caption{Diagram for excited quark production and decay.}
  \label{fig:diagram}
\end{figure}

Given the rich physics potential of multi-jet events~\cite{atlasmj,cmsmj} produced from LHC proton-proton collision, in this letter we perform an analysis of LHC data to search for excited quark production.  We rely on the cases in which excited quarks are produced through contact interaction and then decay into quarks and gluons through gauge interaction, leading to a multi-jet signature (see Fig.~\ref{fig:diagram}).

Our results are derived from the recent ATLAS search~\cite{atlasmj} in final states with multiple high-transverse-momentum jets based on 20.3 fb$^{-1}$ of integrated luminosity, which was originally designed to search for low-scale gravity models. We interpret our results in the context of excited quark production through a contact interaction which produces an excited quark in association with a quark. The excited quark then decays into a quark and a gluon. The final state therefore contains at least three energetic jets. 

Events are selected with~\cite{atlasmj}:
\begin{itemize}
\item At least three AntiKt4 jets with $p_{T}>50$~GeV and $|\eta|<2.8$;
\item Event $H_T$ (scalar sum of jet transverse momenta) at least 1.5 TeV.
\end{itemize}

\begin{figure}[htbp]
  \includegraphics[width=0.9\linewidth]{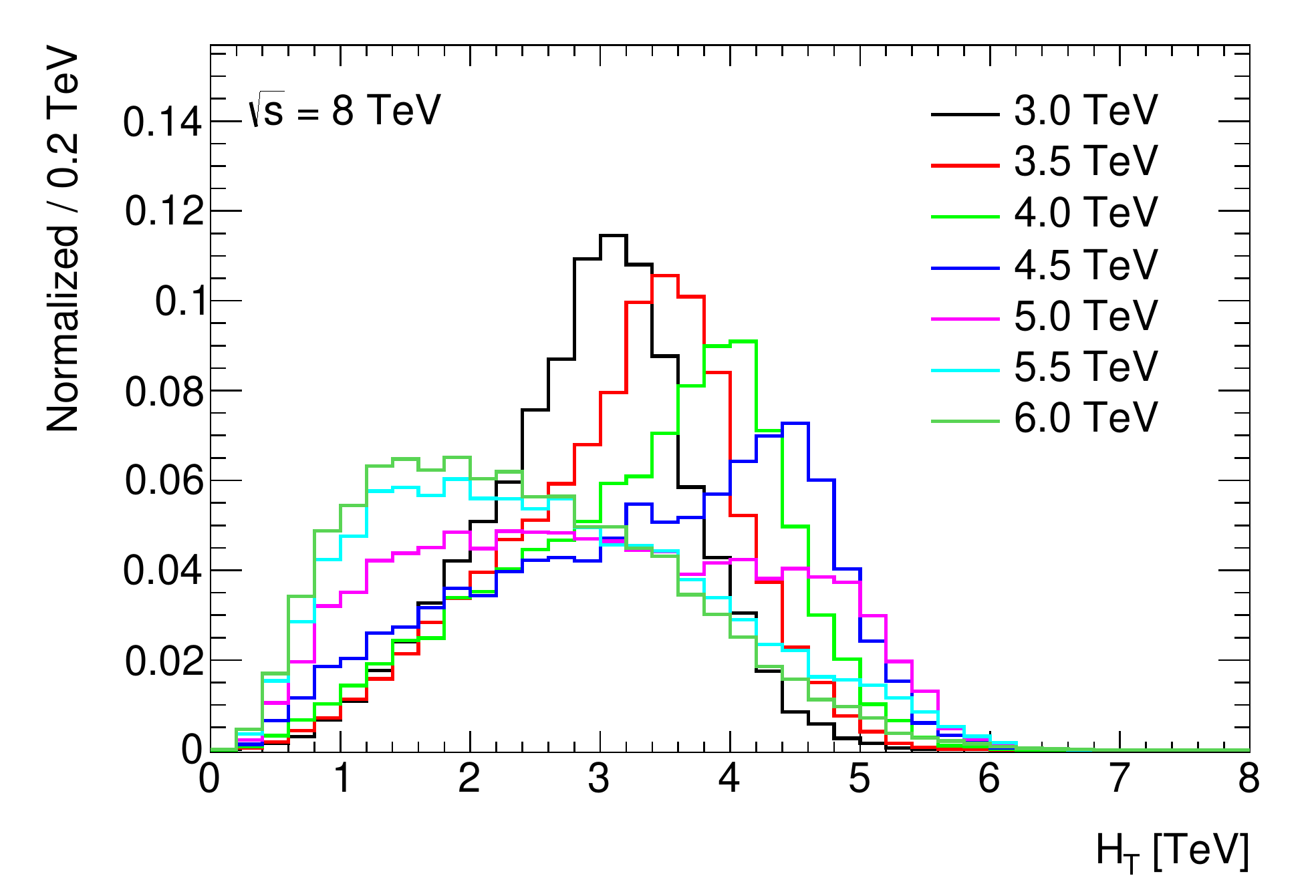}
  \includegraphics[width=0.9\linewidth]{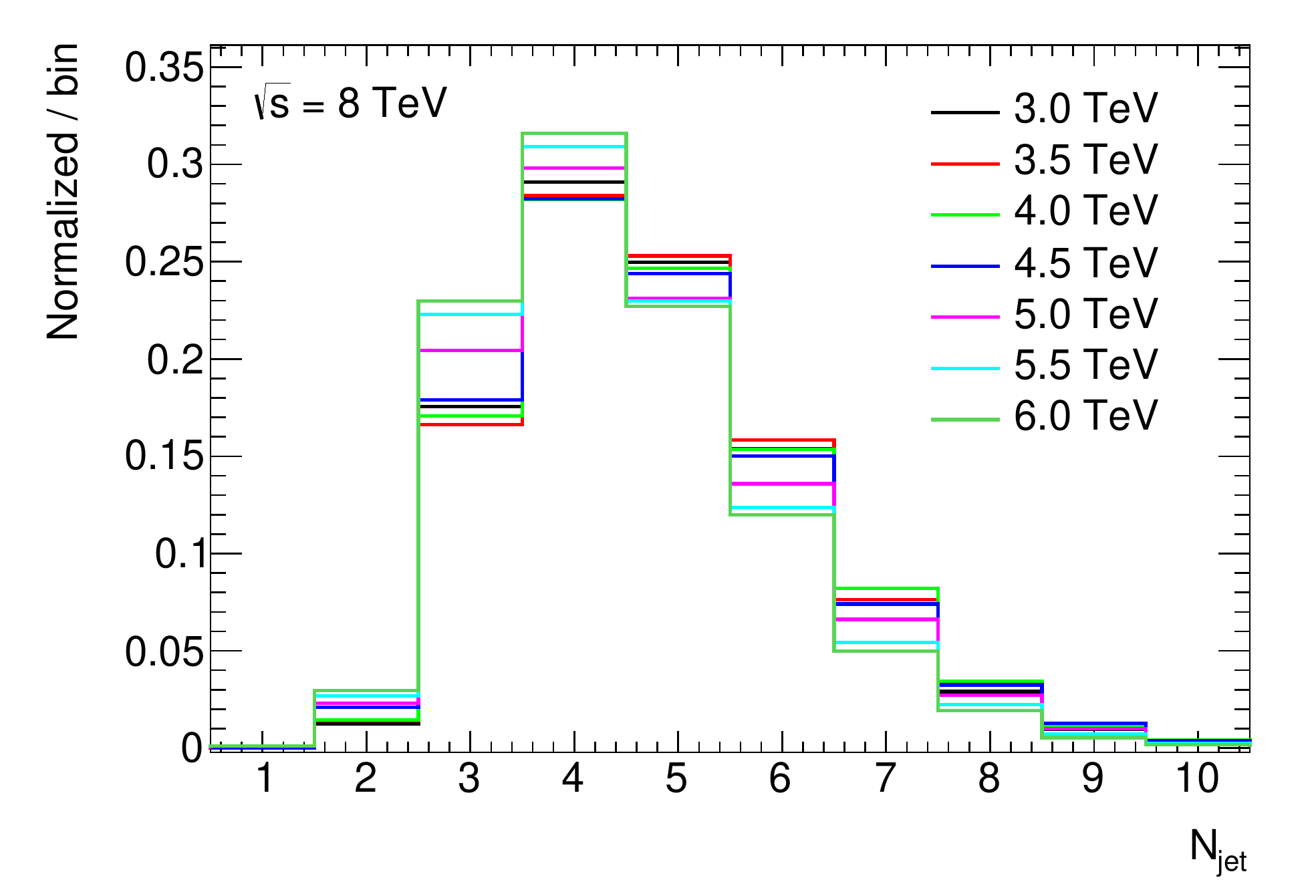}
  \includegraphics[width=0.9\linewidth]{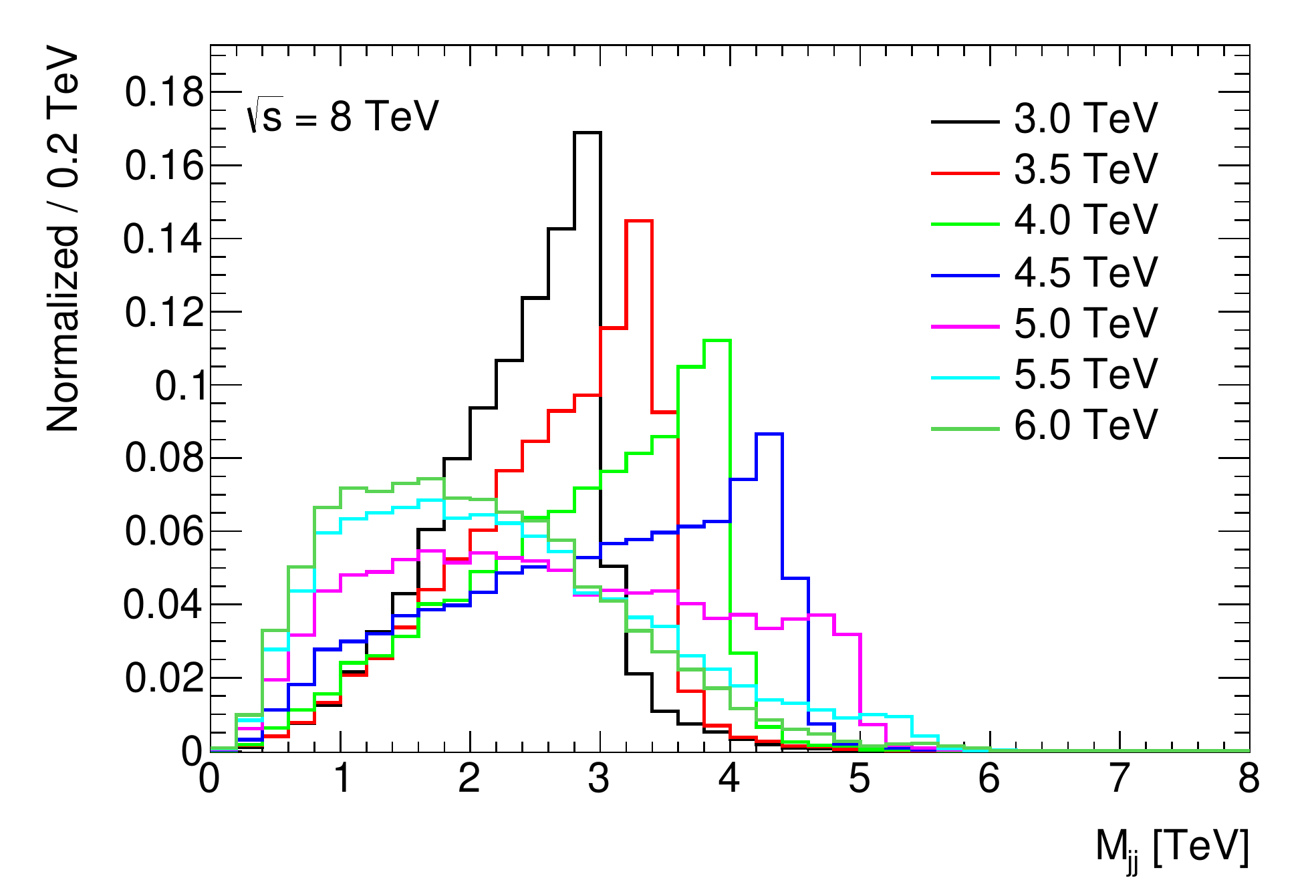}
  \caption{ Normalized distributions of the scalar sum of jet transverse momenta $\rm H_{T}$, jet multiplicity $\rm N_{jet}$ and the invariant mass of leading two jets $\rm M_{jj}$. The solid colored lines correspond to different excited quark masses.}
  \label{fig:kin}
\end{figure}

Six overlapping signal regions of inclusive jet multiplicity (from $N_{jet}\ge 3$ to $N_{jet}\ge 8$) were defined and a scan of multiple inclusive $H_T$ regions ($H_T \ge H_{T}^{min}$) for excess were performed. 
Upper limits on the signal visible cross section, which is the result of signal production rate times the acceptance times the reconstruction efficiency, were calculated through the cut-and-count analysis as a function of $H_{T}^{min}$ for each inclusive jet multiplicity from $N_{jet} \ge 3$ to $N_{jet} \ge 8$.
\[ \sigma_{visible} = \sigma \times A \times \epsilon  \]
\noindent
where $\sigma$ is the production cross section; $A$ is the acceptance, the fraction of produced events satisfying the
event selection; $\epsilon$ is the detector reconstruction efficiency which is $90\%$ in this analysis.

Applying this limit to derive a cross-section limit for an arbitrary process requires knowing the signal acceptance $A$ for the model of interest. 

We simulate excited quark production with showering and hadronization using {\sc pythia}~\cite{pythia}. The detector simulation is applied by a parametric fast simulation tuned to match the ATLAS
performance~\cite{pgs}. In our analysis, the quark compositeness scale $\Lambda$ is chosen as the excited quark mass $m_{q^{*}}$; all dimensionless constants $f_{s}$, $f$, and $f'$ are assumed to be 1.
Only the first generation excited quark state ($u^{*}$ and $d^{*}$) are considered in this letter.

\begin{figure}[htpb]
\includegraphics[width=0.49\linewidth]{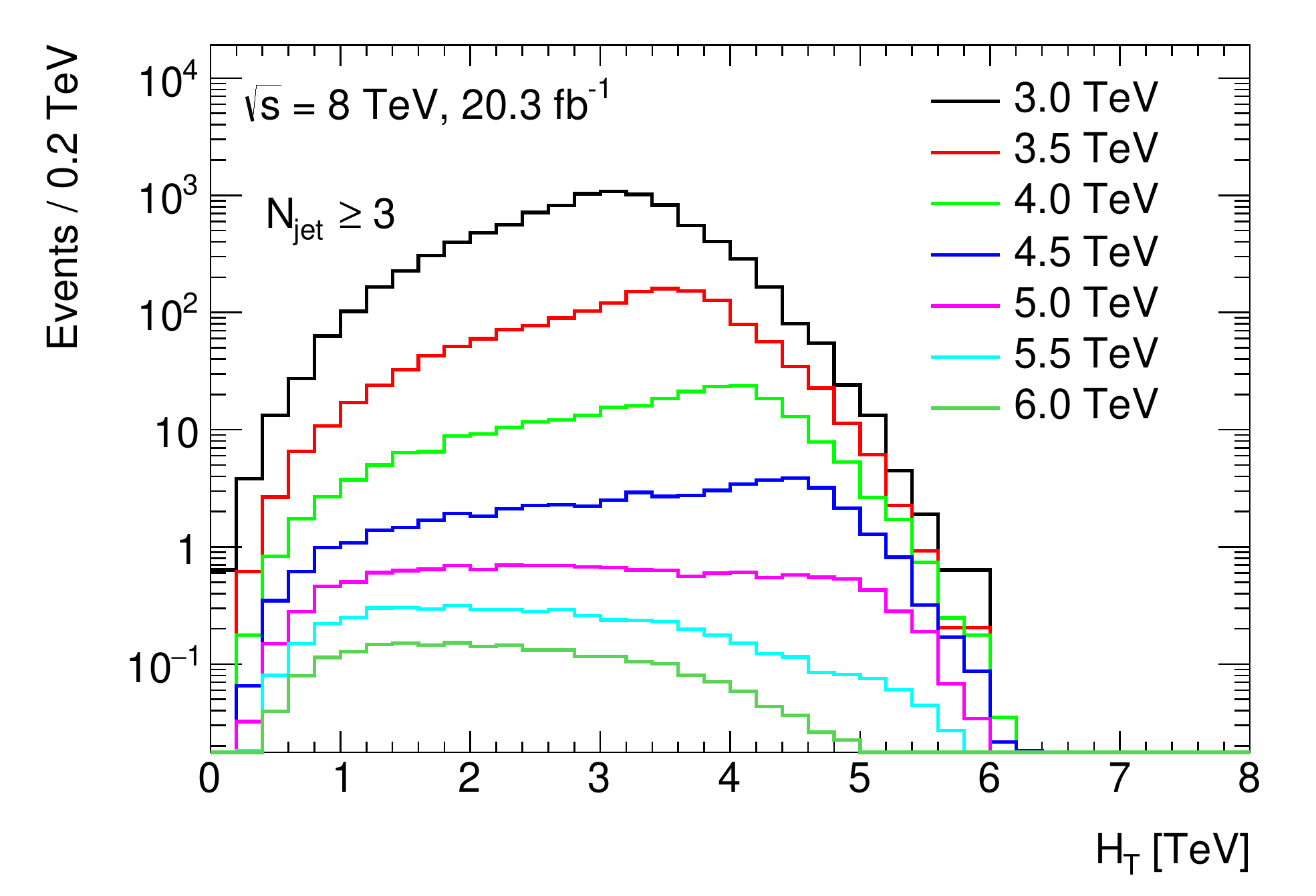}
\includegraphics[width=0.49\linewidth]{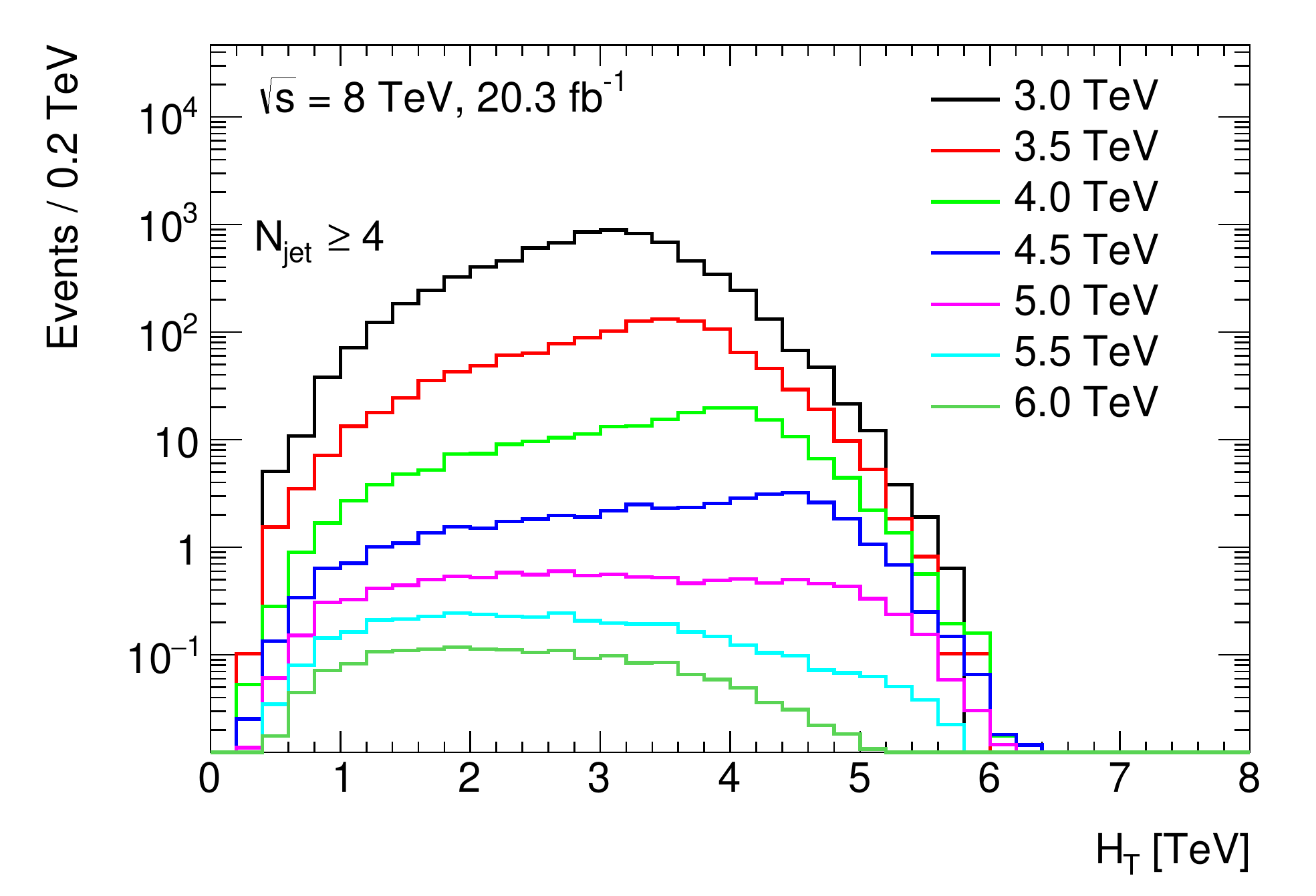}
\includegraphics[width=0.49\linewidth]{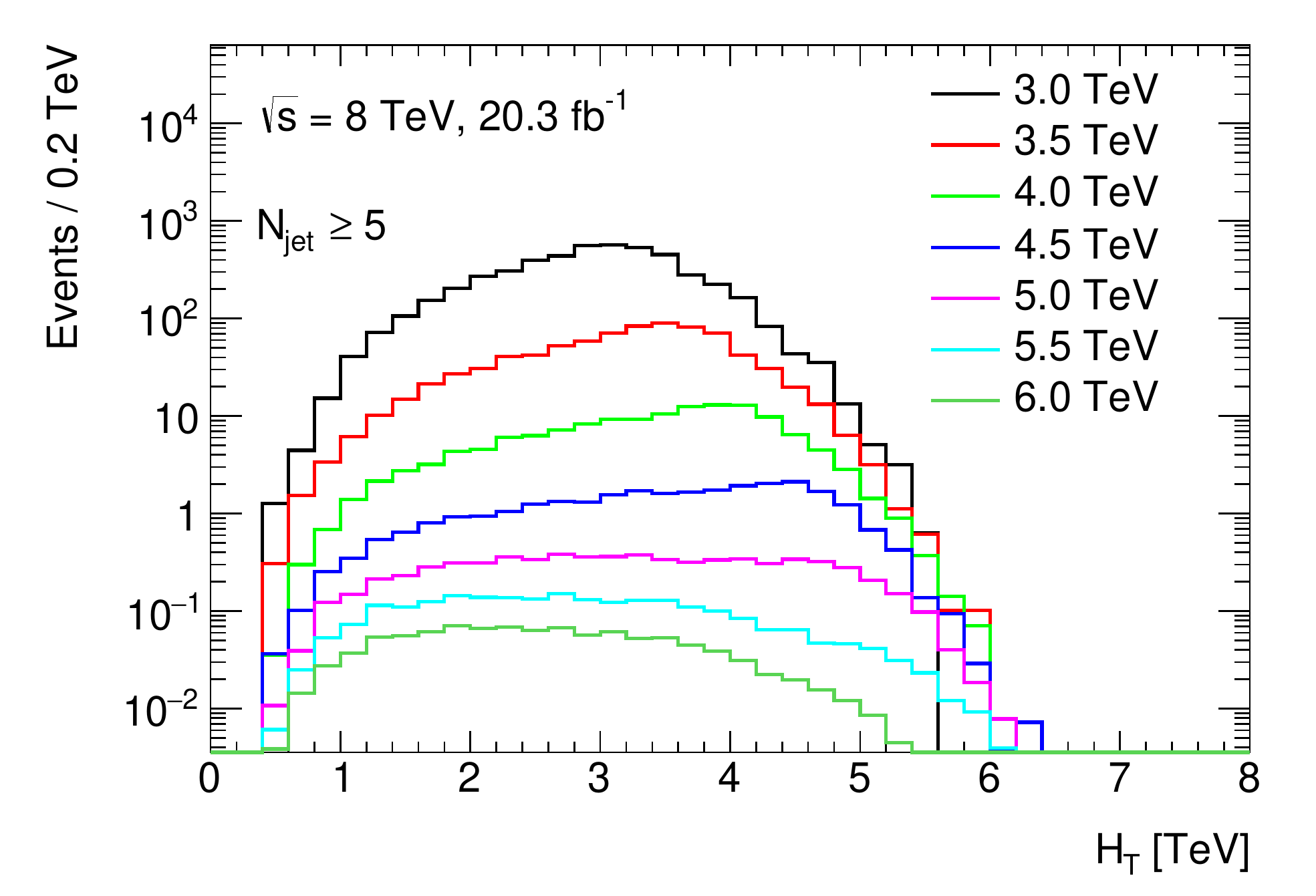}
\includegraphics[width=0.49\linewidth]{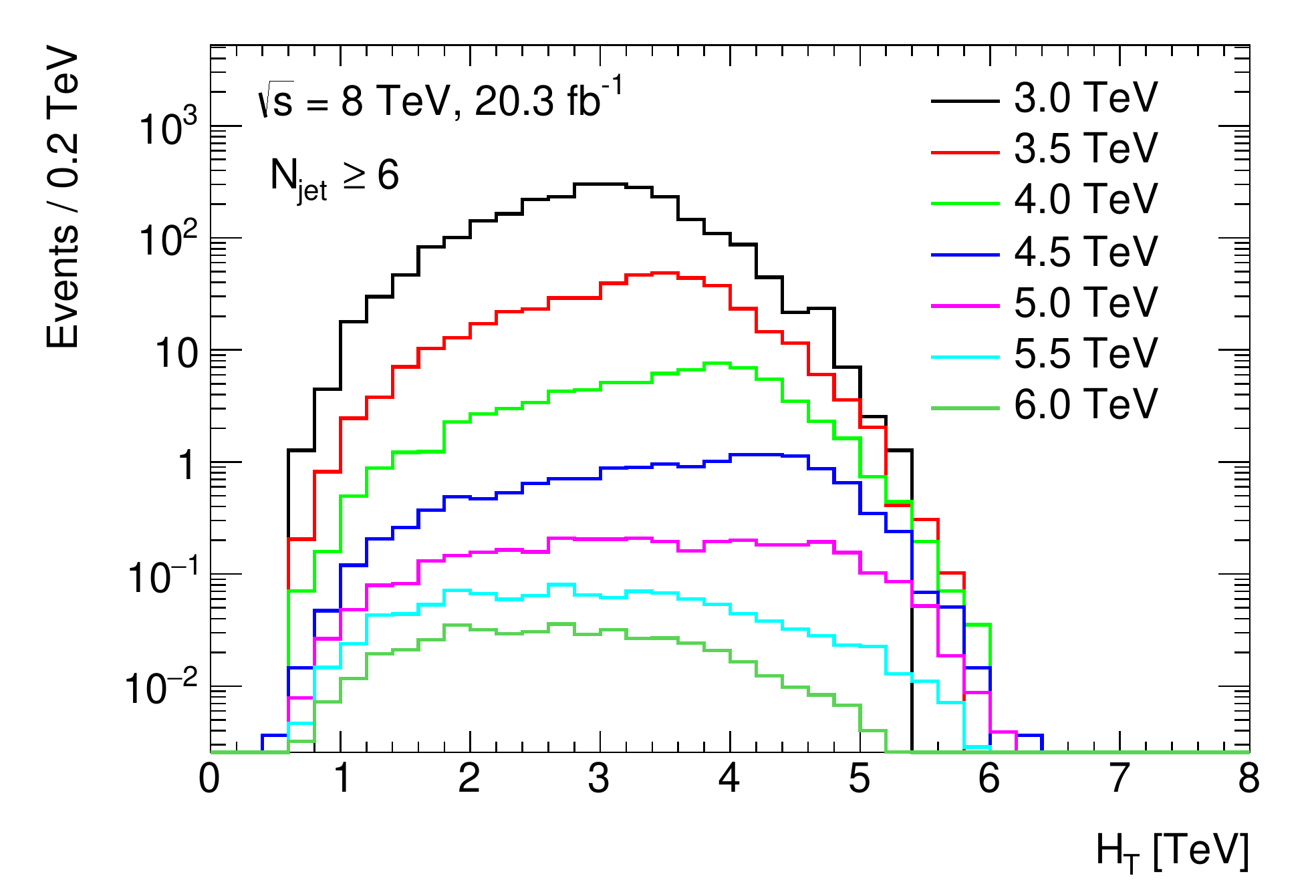}
\includegraphics[width=0.49\linewidth]{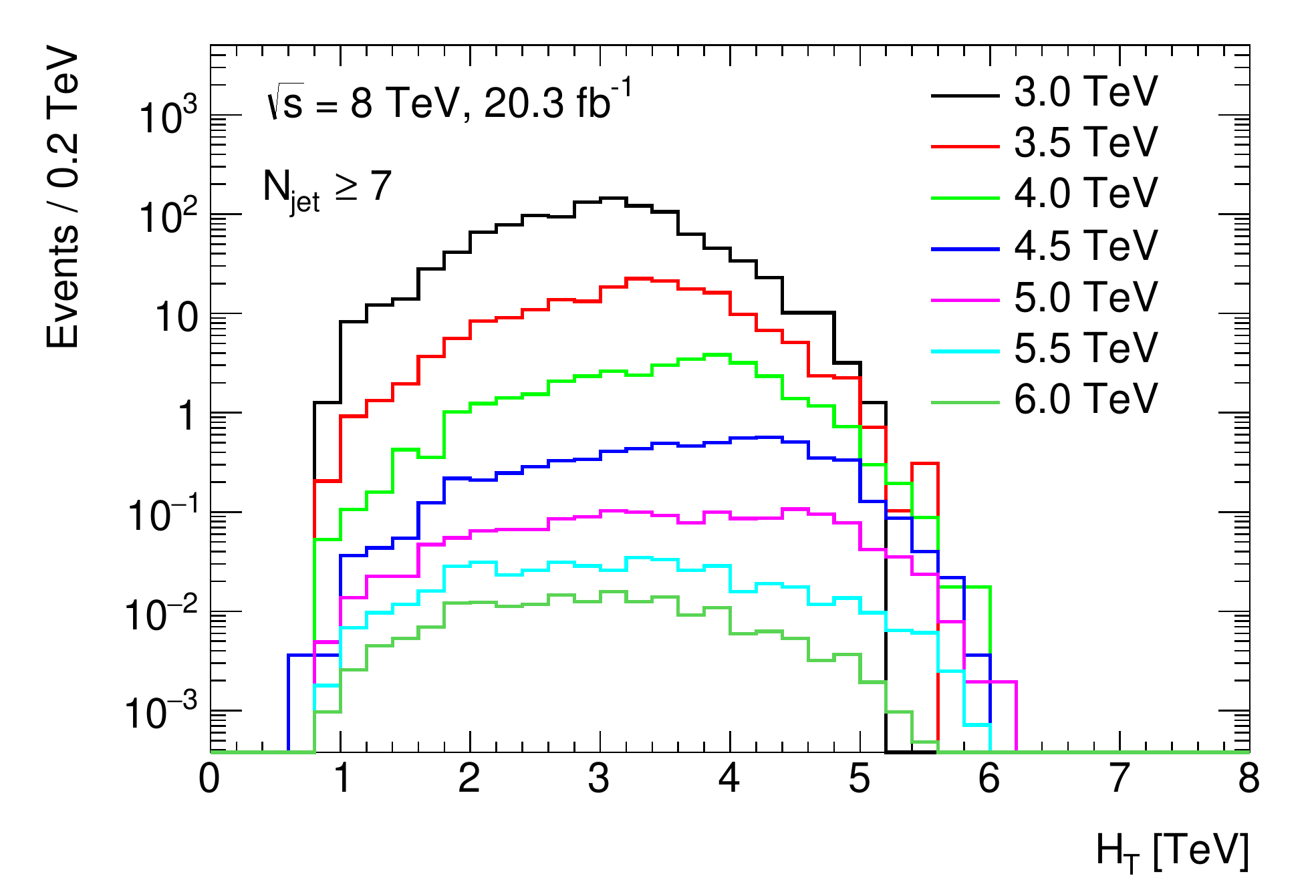}
\includegraphics[width=0.49\linewidth]{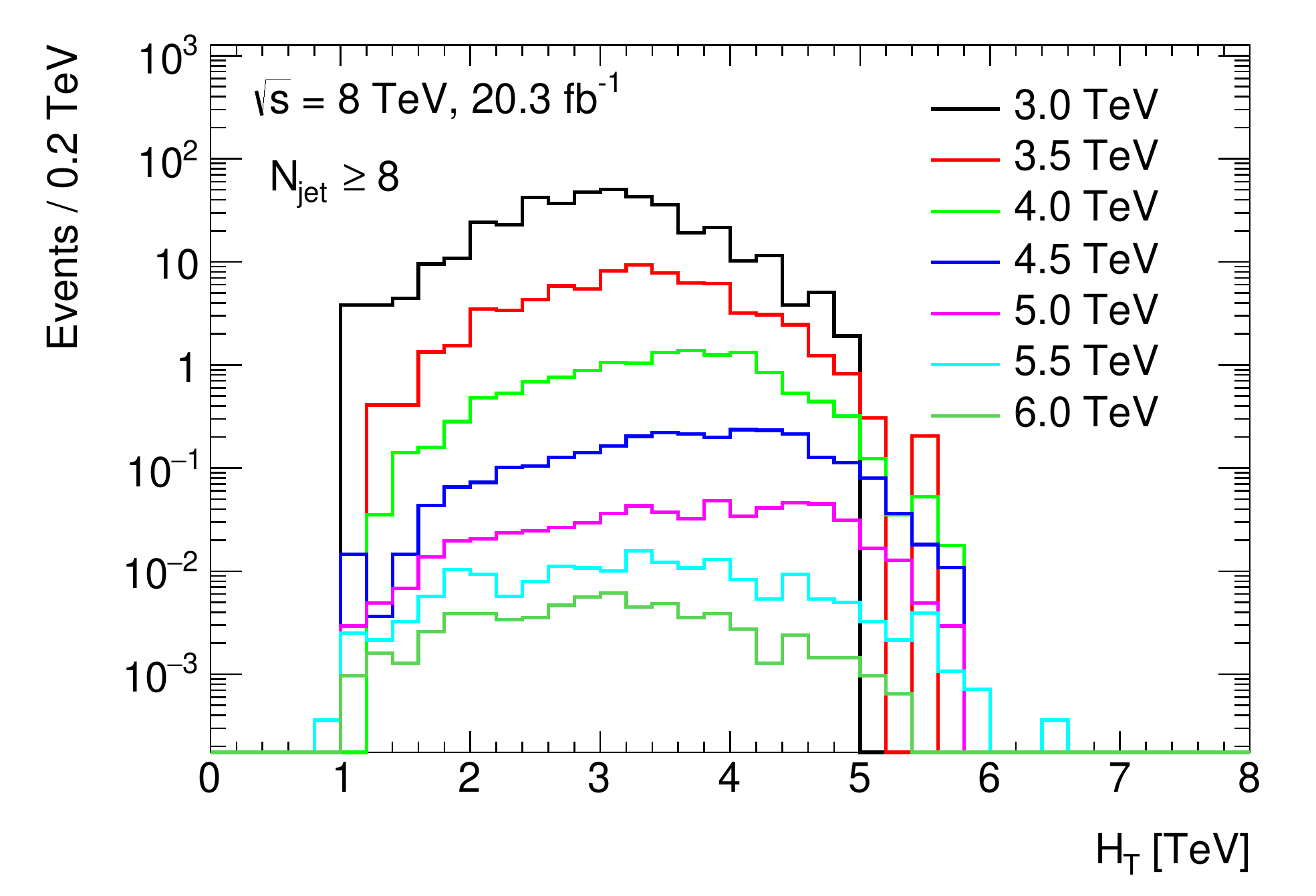}
\caption{Distributions of the scalar sum of jet transverse momenta $\rm H_{T}$ for different jet multiplicities. The solid colored lines correspond to different excited quark masses.}
  \label{fig:kin2}
\end{figure}

Some representative kinematic distributions are shown in Fig.~\ref{fig:kin} and Fig.~\ref{fig:kin2}.

\begin{table}[htpb]
\center
{
\begin{tabular}{lr}
\hline\hline
& $N_{jet}\ge 4$, $H_T \ge 4.1$~TeV \\ \hline
\vspace{2mm}
Data & 0 \\
\vspace{2mm}
Background & $2.034 ^{+1.053}_{-1.598}$ \\
\vspace{2mm}
Expected limit $\sigma_{visible}$ & $0.145 ^{+0.073}_{-0.004}$ fb \\
\vspace{2mm}
Observed limit $\sigma_{visible}$ & 0.142 fb \\ \hline \hline
\end{tabular}
}
\caption{Data yields and derived visible cross section limits from ATLAS multi-jet search~\cite{atlasmj}.}
\label{tab:yields}
\end{table}

\begin{figure}[htpb]
  \center
  \includegraphics[width=0.49\linewidth]{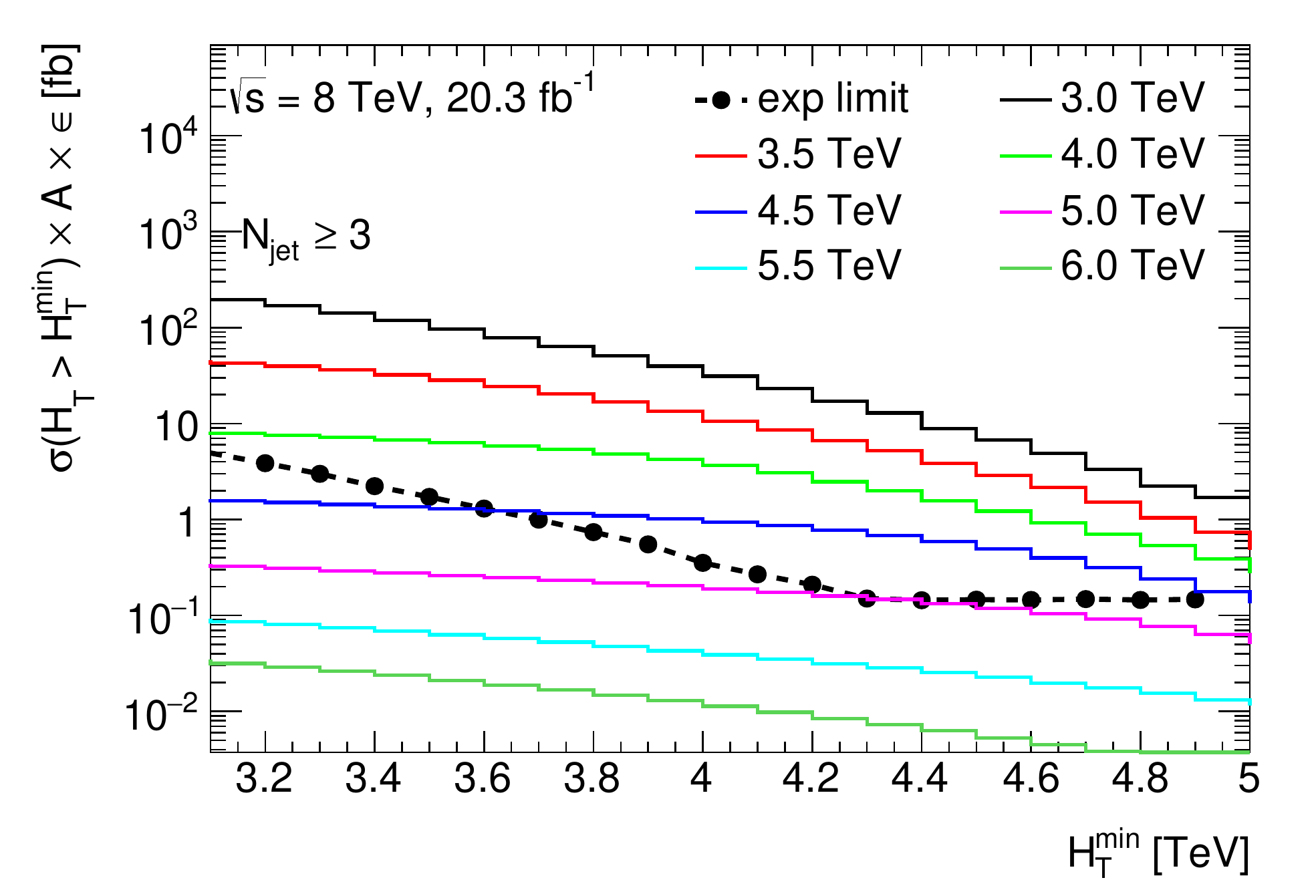}
  \includegraphics[width=0.49\linewidth]{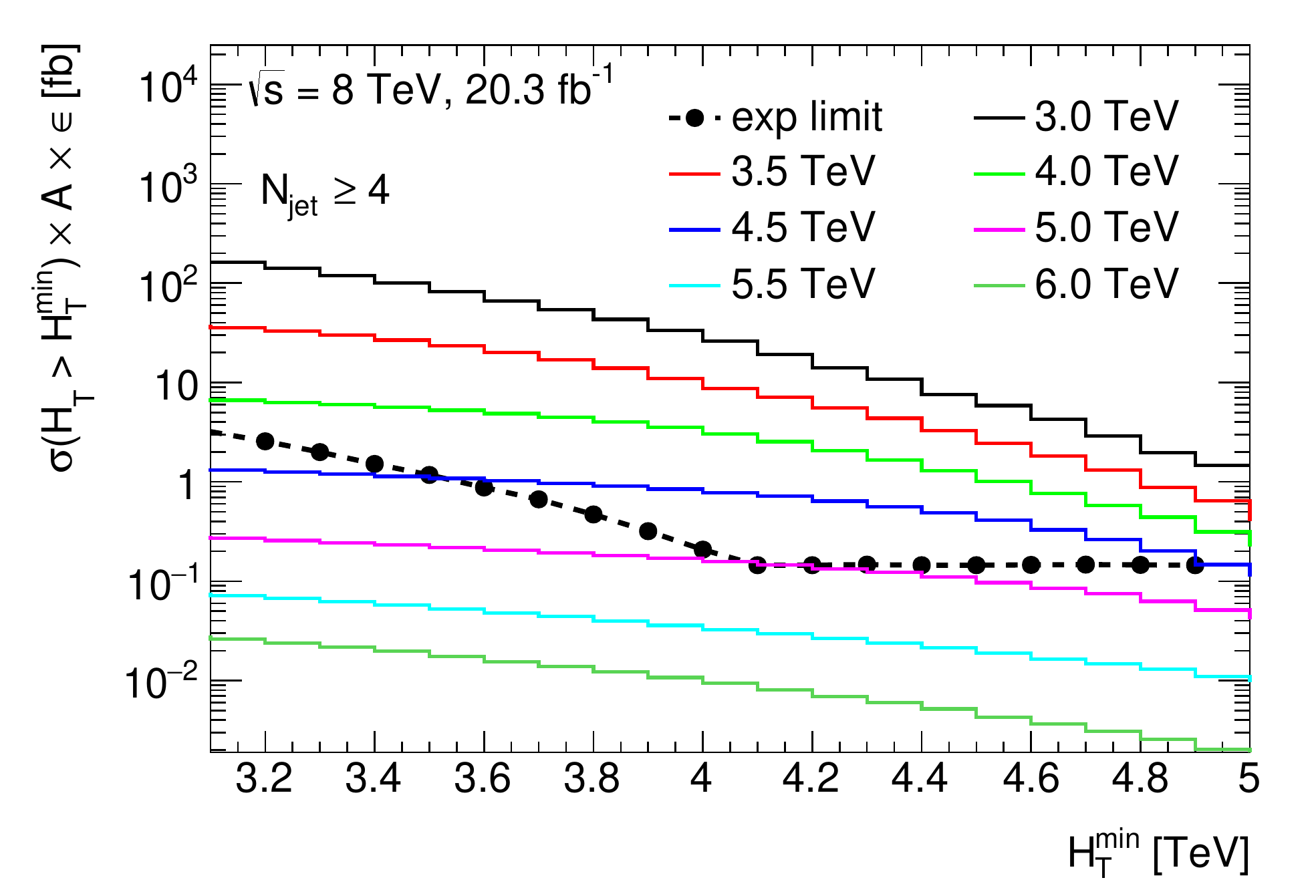}
  \includegraphics[width=0.49\linewidth]{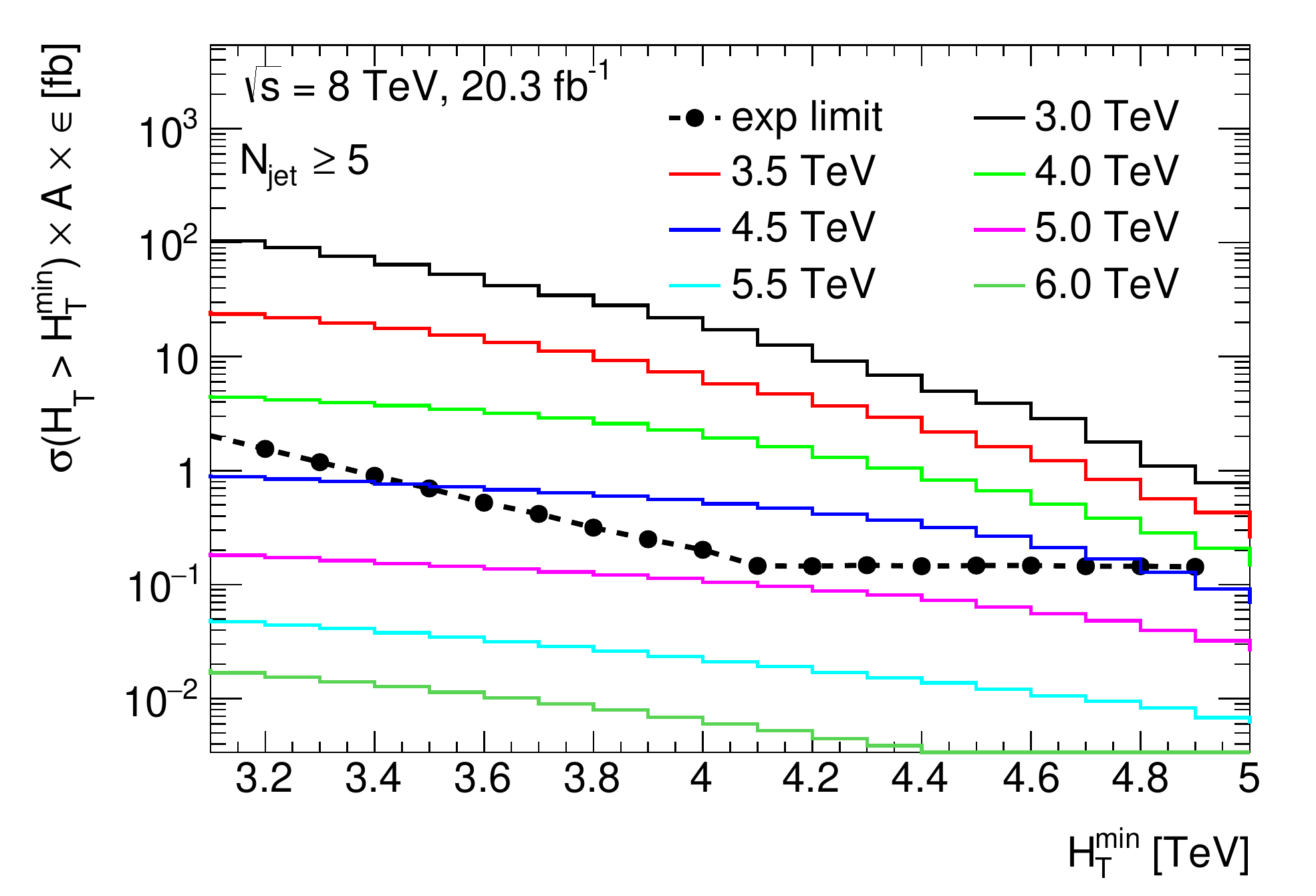}
  \includegraphics[width=0.49\linewidth]{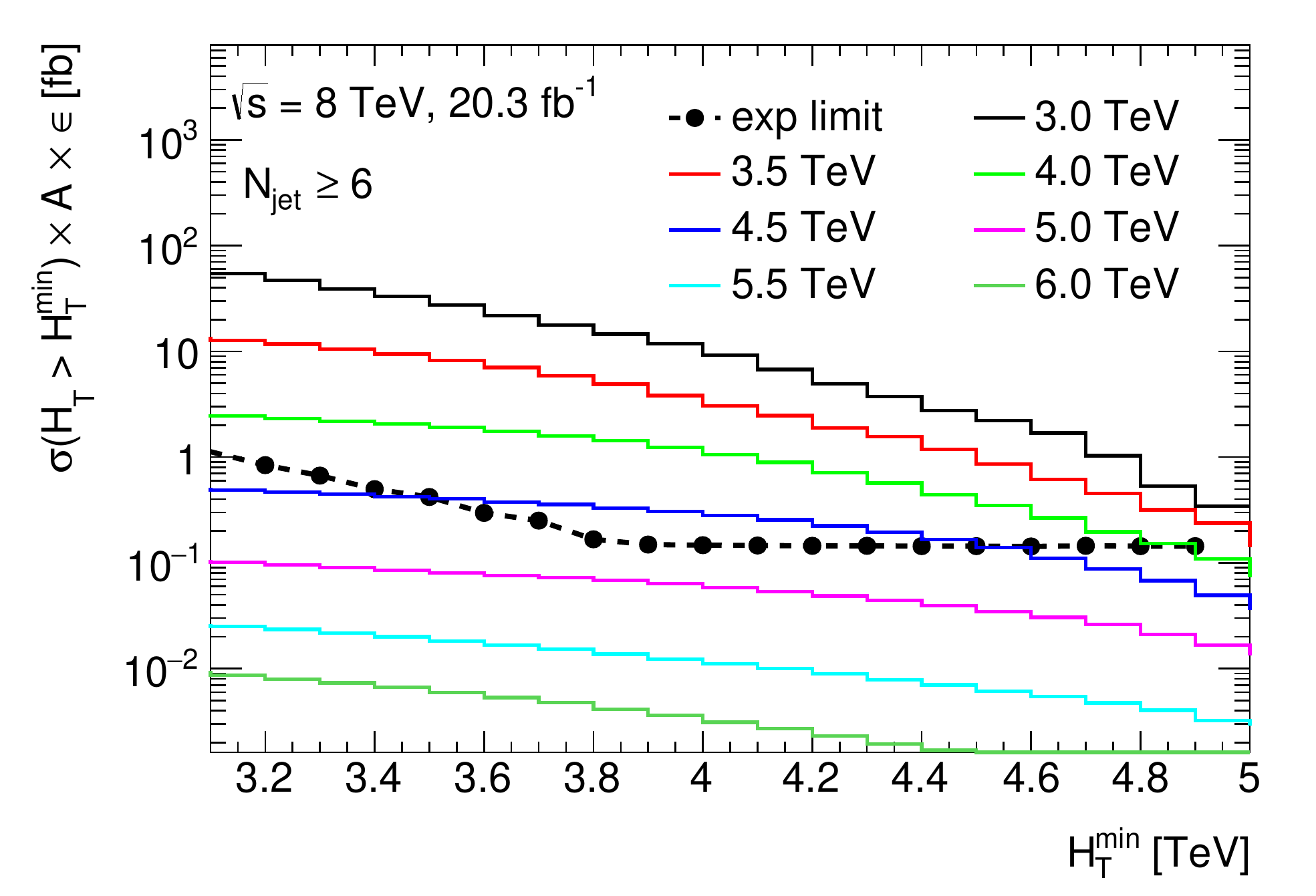}
  \includegraphics[width=0.49\linewidth]{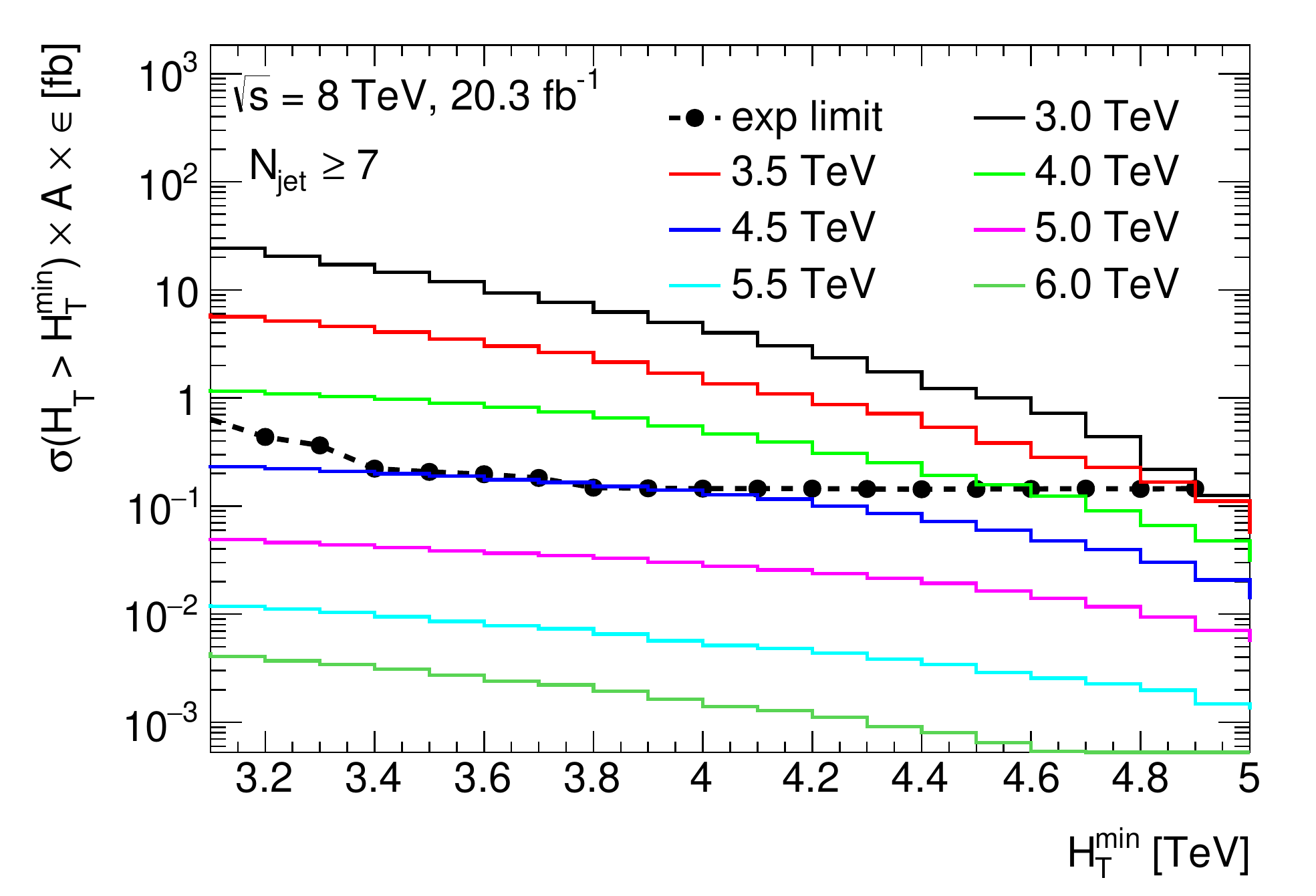}
  \includegraphics[width=0.49\linewidth]{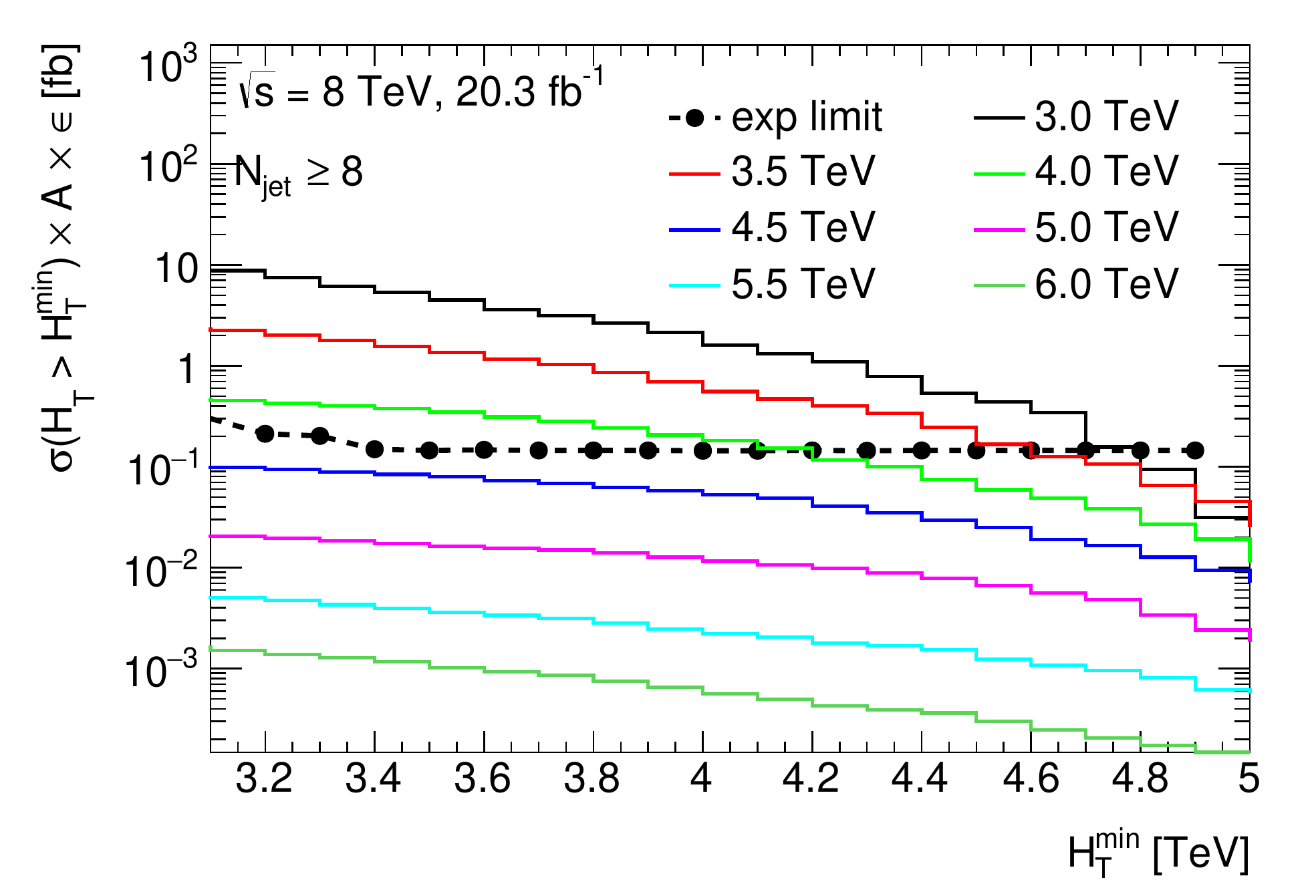}
  \caption{ Signal visible cross section as a function of $H_{T}^{min}$ for different jet
    multiplicities. The solid colored lines correspond to different excited quark masses. The expected limit on visible cross section from ATLAS multi-jet search is shown in black dashed line with dot markers.}
  \label{fig:compare}
\end{figure}

In order to find the optimal jet multiplicity and $H_T$ cuts for the excited quark signal, we compare the signal visible cross section with the expected exclusion limit, as shown in Fig.~\ref{fig:compare}.

From this comparison, signal region with $N_{jet}\ge 4$ and $H_{T}\ge 4.1$~TeV can give the best sensitivity for excited quark on average. The corresponding number of selected events and the estimates of the SM background contributions are shown in Table~\ref{tab:yields}~\cite{atlasmj}.

\begin{figure}[htbp]
\includegraphics[width=0.9\linewidth]{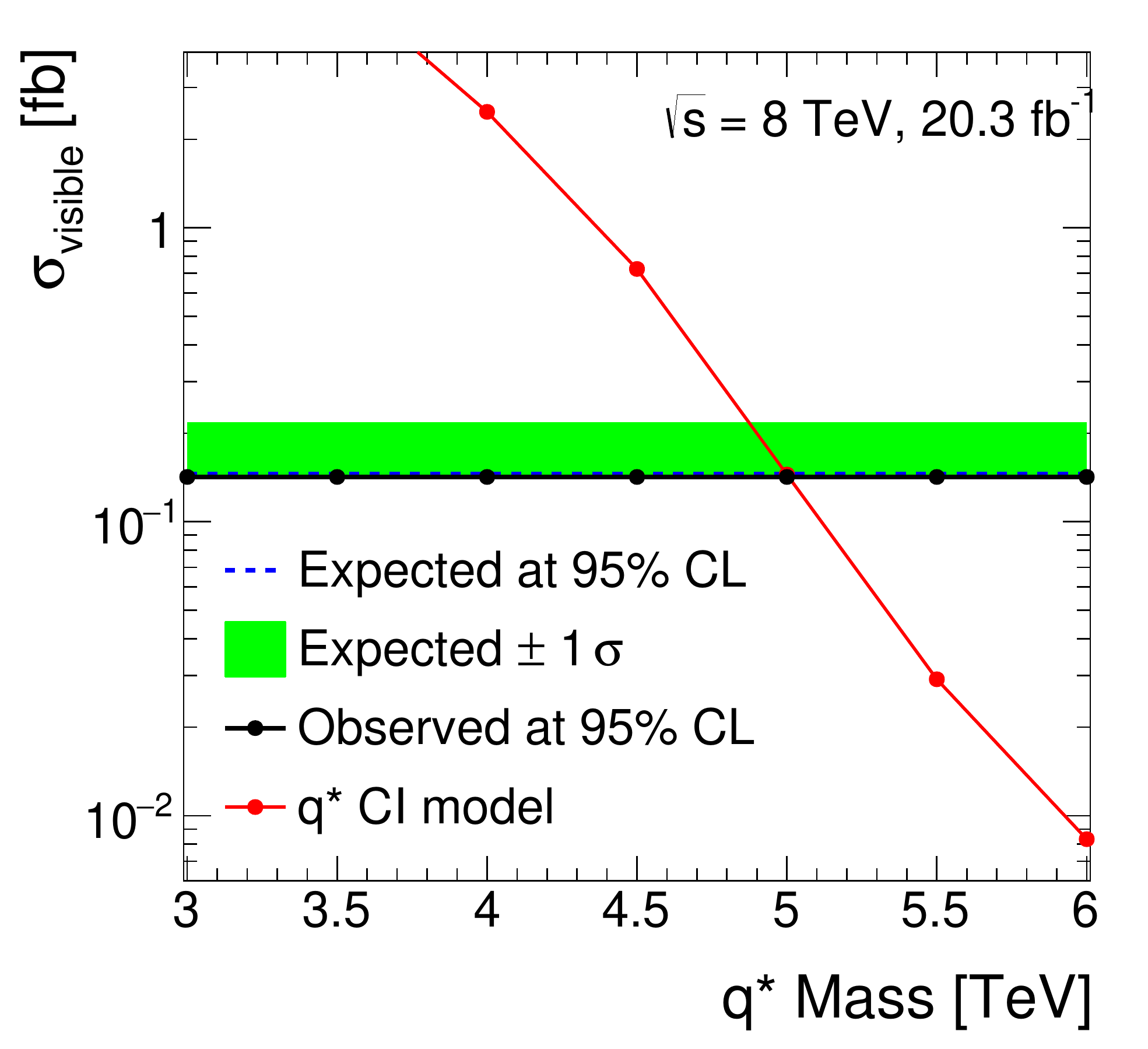}
\caption{Exclusion limits on the visible cross section for various excited quark signals without considering the model validity.}
\label{fig:limit}
\end{figure}

In this signal region, the acceptance of various excited quark signals is listed in Table~\ref{tab:acc}. With these acceptance values, the visible cross section limits are compared to the excited quark theoretical cross section, as shown in Fig~\ref{fig:limit}. It indicates that excited quark masses below 5~TeV are excluded by the ATLAS multi-jet search at $95\%$ confidence level.

\begin{table}[h]
\center
{
\begin{tabular}{lccccccc}
\hline \hline
$m_{q^{*}}$  & 3~TeV & 3.5~TeV & 4~TeV & 4.5~TeV & 5~TeV & 5.5~TeV & 6.0~TeV\\
$A$ & 0.04 & 0.09 & 0.19 & 0.27 & 0.20 & 0.11 & 0.07 \\
\hline \hline
\end{tabular}
}
\caption{Acceptance of various excited quark signals.}
\label{tab:acc}
\end{table}

\begin{figure}[htpb]
  \center
  \includegraphics[width=0.49\linewidth]{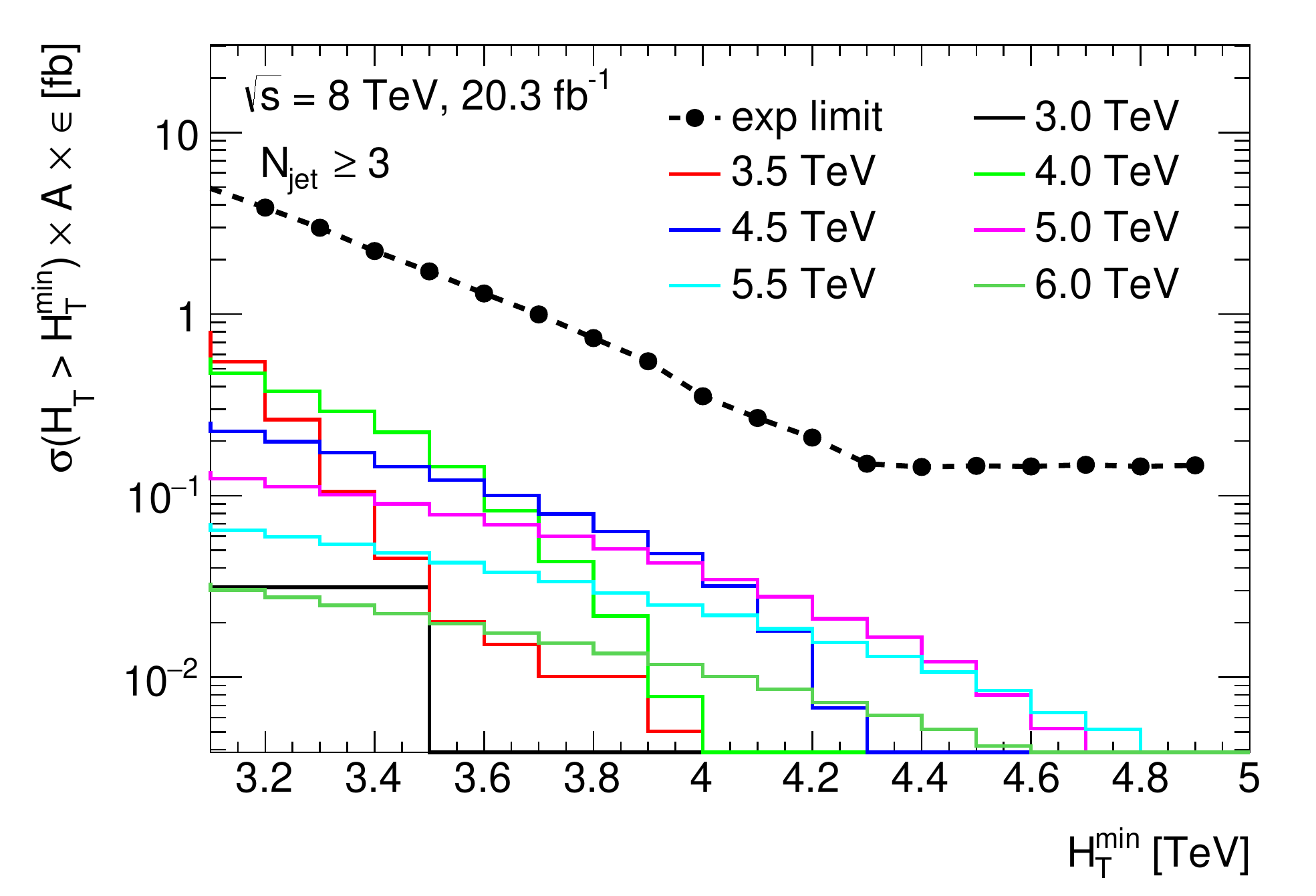}
  \includegraphics[width=0.49\linewidth]{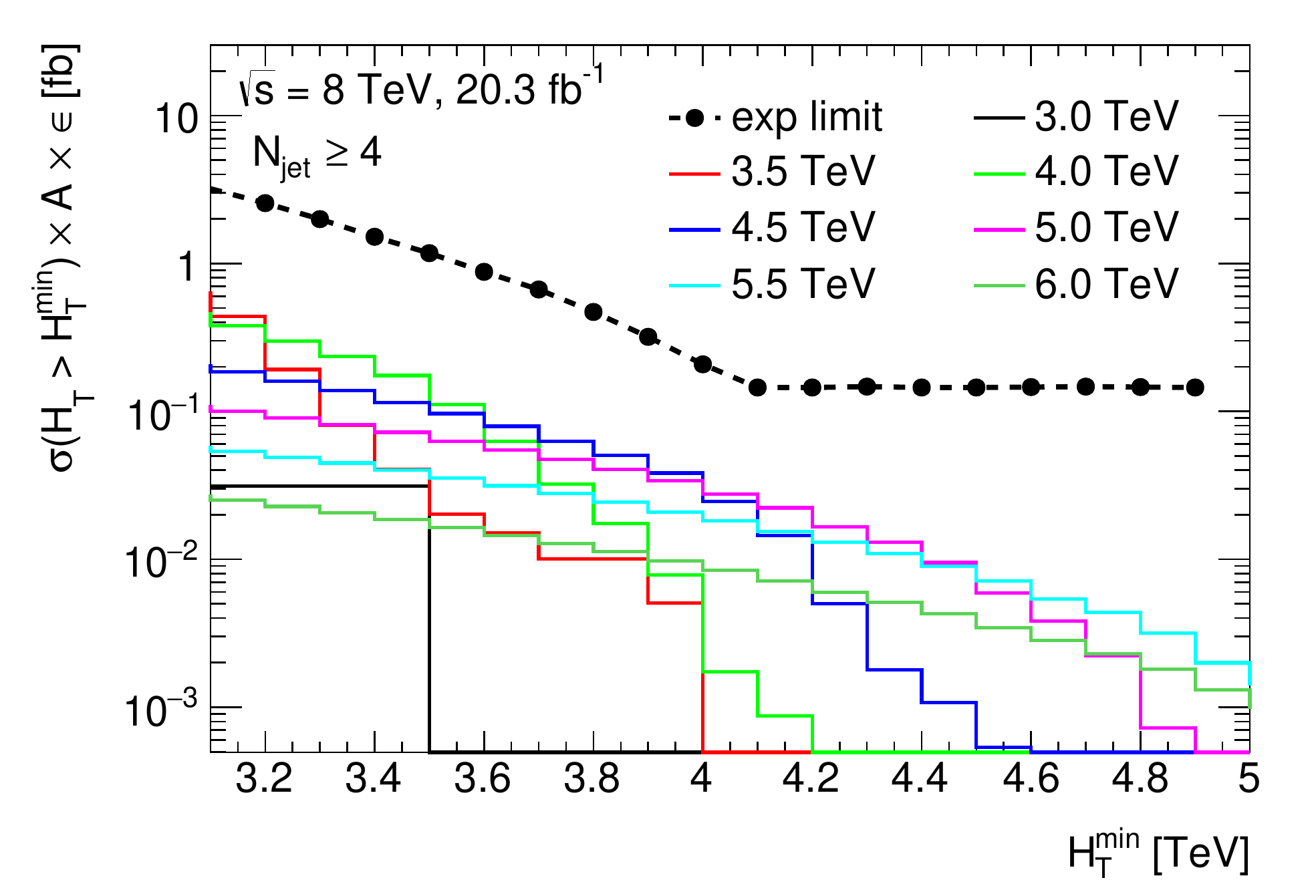}
  \includegraphics[width=0.49\linewidth]{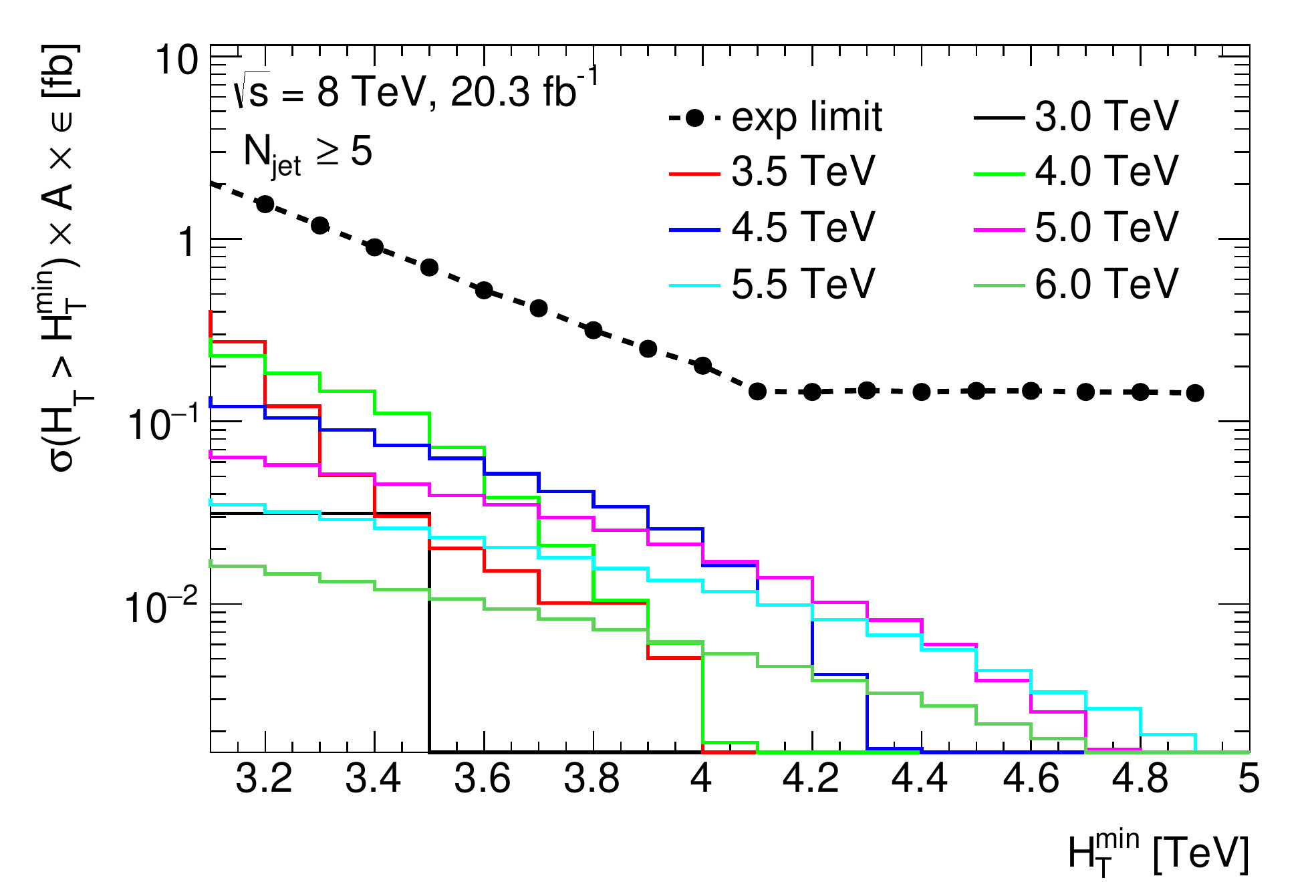}
  \includegraphics[width=0.49\linewidth]{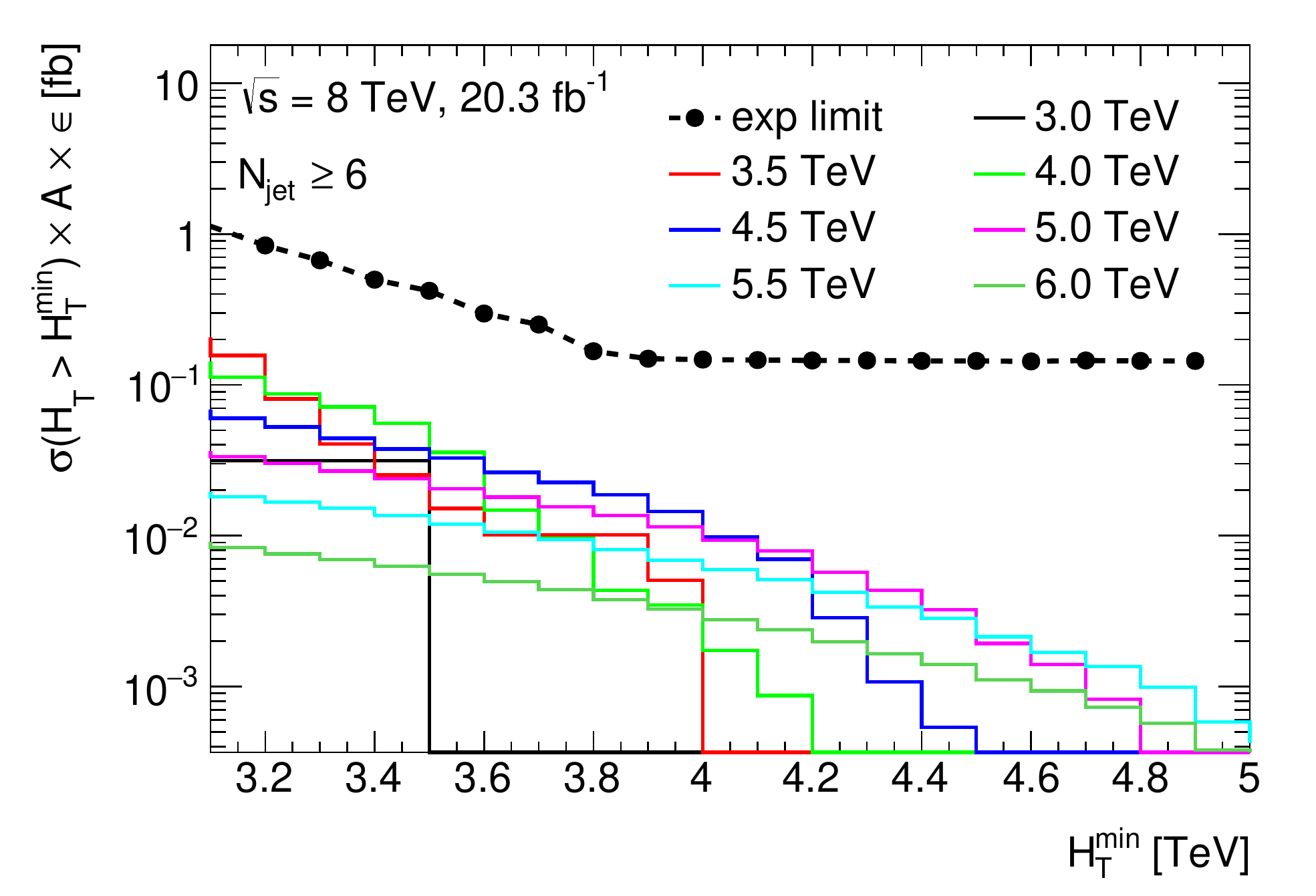}
  \includegraphics[width=0.49\linewidth]{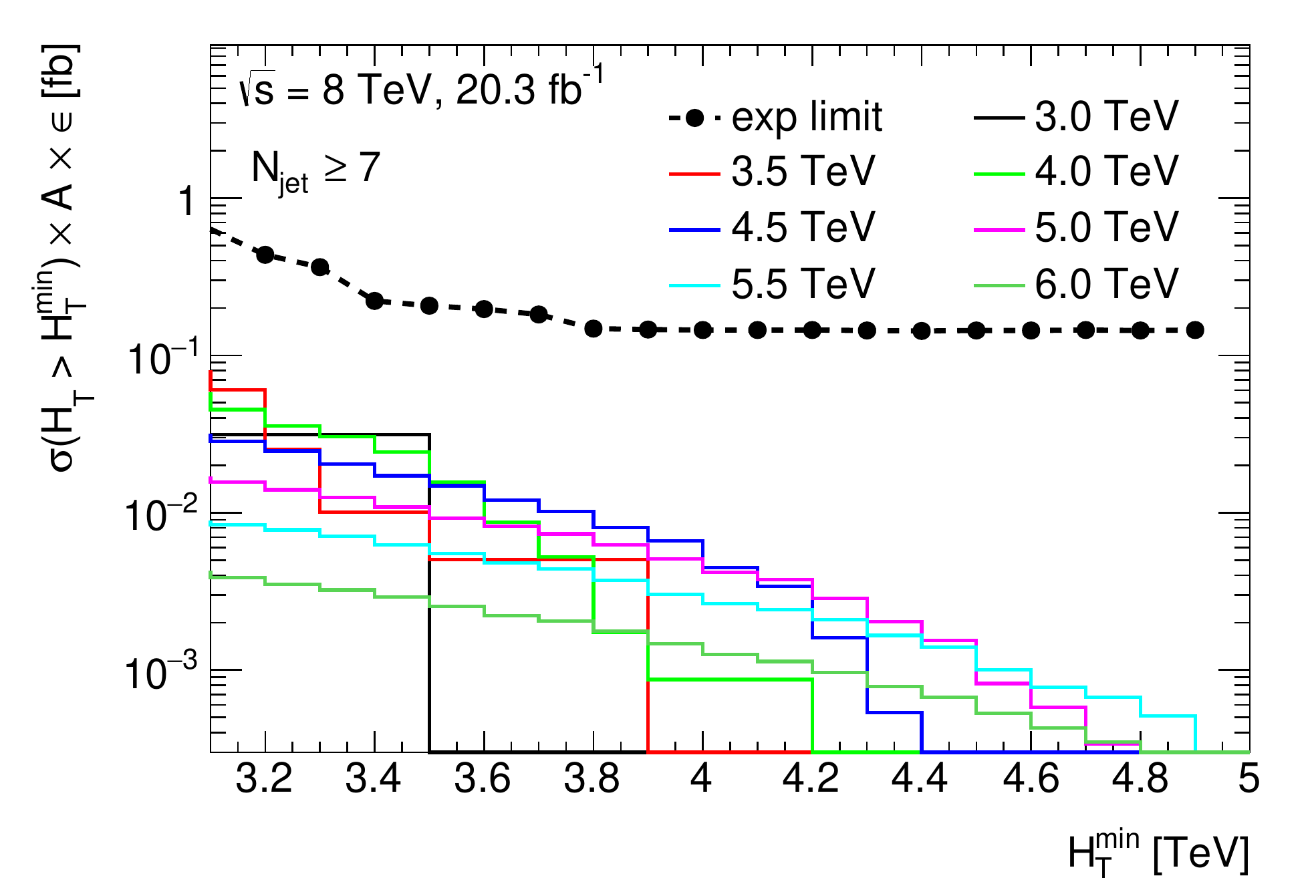}
  \includegraphics[width=0.49\linewidth]{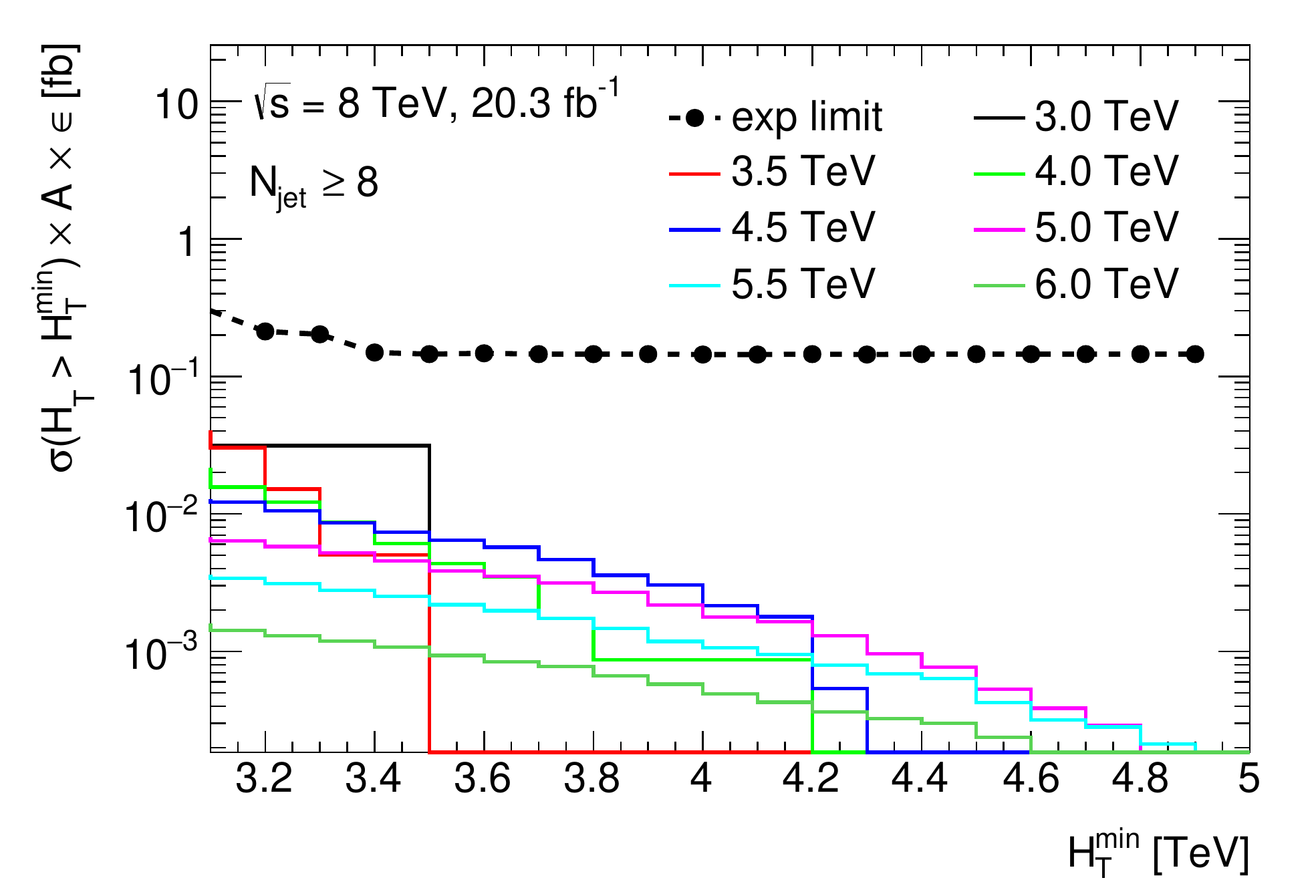}
  \caption{ Signal visible cross section as a function of $H_{T}^{min}$ for different jet
    multiplicities after the validity truncation. The solid colored lines correspond to different excited quark masses. The expected limit on visible cross section from ATLAS multi-jet search is shown in black dashed line with dot markers.}
  \label{fig:valid}
\end{figure}

Since the EFT frame is valid only when the energy scale of the new physics is much larger compared with the center of mass energy of the collider, the EFT model becomes a poor approximation when the momentum transferred in the interaction, $Q_{tr}$, is comparable to the compositeness scale $\Lambda$. In order to illustrate the sensitivity to the unknown ultraviolet completion of the theory, we computed the limits retaining only simulated events with $Q_{tr}<\Lambda$ (truncation). After the validity truncation, the remaining signal visible cross section is shown in Fig.~\ref{fig:valid}, which drops a lot and turns to be lower than current expected limit. It indicates the importance of considering the validity of EFT model.

In closing, we have set bounds on the excited quark production via contact interaction, using the recent ATLAS multi-jet search.  While this search is effective in placing interesting limits on excited quark production, it uses the information of scalar sum of jet transverse momenta only. One can improve it by applying more kinematic distributions, like the invariant mass of two jets which might be reconstructed as the excited quark mass. When EFT validity is concerned, a large fraction of signal phase space is removed which largely compromises the collider constraints on this excited quark model. Our work is only a first step into what should be a fruitful age of excited quark searches at the LHC.

\subsection{Acknowledgements}
The authors are grateful for the useful conversations with Jun Guo, Qing Wang, Zuowei Liu and Xin Chen. We acknowledge the support from Tsinghua University, Center of High Energy Physics of Tsinghua University and Collaborative Innovation Center of Quantum Matter of China. Kechen Wang is supported in part by the CAS Center for Excellence in Particle Physics (CCEPP) and wants to thank Cai-Dian L$\rm \ddot{u}$ for his help.


\begin{thebibliography}{99}
\bibitem{Pati:1974}
  J.C.~Pati and A.~Salam, Phys.\ Rev.\ D{\bf 10}, 275 (1974)
\bibitem{Pati:1975}
  J.C.~Pati and A.~Salam, Phys.\ Rev.\ D{\bf 11}, 703 (1975)
\bibitem{Eichten:1983} 
  E.J.~Eichten, K.D.~Lane and M.E.~Peskin,
  Phys.\ Rev.\ Lett. {\bf 50}, 811 (1983)
\bibitem{Cabibbo:1984}
  N.~Cabibbo, L.~Maiani and Y.~Srivastava,
  Phys.\ Lett.\ B {\bf 139}, 459 (1984)
\bibitem{Hagiwara:1985}
  K.~Hagiwara, D.~Zeppenfeld and S.~Komamiya,
  Z.\ Phys.\ C {\bf 29}, 115 (1985)
\bibitem{Baur:1987}
  U.~Baur, I.~Hinchliffe, D.~Zeppenfeld, Int.\ J.\ Mod.\ Phys.\ A{\bf 2}, 1285(1987)
\bibitem{Baur:1990}
  U.~Baur, M.~Spira and P.M.~Zerwas,
  Phys.\ Rev.\ D {\bf 42}, 815 (1990)
\bibitem{Redi:2013}
  M.~Redi, V.~Sanz, M.~de Vries, A.~Weiler,
  JHEP\ 08(2013) 008
\bibitem{Bhattacharya:2009xg} 
  S.~Bhattacharya, S.~S.~Chauhan, B.~C.~Choudhary and D.~Choudhury,
  Phys.\ Rev.\ D {\bf 80}, 015014 (2009)
  doi:10.1103/PhysRevD.80.015014
  [arXiv:0901.3927 [hep-ph]].

\bibitem{pdg:2012}
  Particle Data Group, Phys.\ Rev.\ D {\bf 86}, 010001 (2012)
\bibitem{hera:2009}
  H1 Collaboration, Phys.\ Lett.\ B {\bf 678}, 335 (2009)
\bibitem{d0:2004}
  D0 Collaboration, Phys.\ Rev.\ D {\bf 69}, 111101 (2004)
\bibitem{cdf:2009}
  CDF Collaboration, Phys.\ Rev.\ D {\bf 79}, 112002 (2009)
\bibitem{cms:2013}
  CMS Collaboration, Phys.\ Rev.\ D {\bf 87}, 114015 (2013)
\bibitem{atlas:2014}
  ATLAS Collaboration, Phys.\ Lett.\ B{\bf 738}, 274-293 (2014)
\bibitem{atlas:2015}
  ATLAS Collaboration, Phys.\ Rev.\ D {\bf 91}, 052007 (2015)
\bibitem{atlasmj}
  ATLAS Collaboration,
  JHEP\ 07(2015) 032
\bibitem{cmsmj}
  CMS Collaboration,
  JHEP\ 07(2013) 178
\bibitem{pythia}
  T.~Sjostrand, S.~Mrenna and P.~Z.~Skands,
  JHEP {\bf 0605}, 026 (2006)
  [hep-ph/0603175].

\bibitem{pgs}
{\tt http://www.physics.ucdavis.edu/~conway/research/\\
software/pgs/pgs4-general.htm}

\end{thebibliography}
\end{document}